\documentclass[]{spie}  

\usepackage{booktabs}
\usepackage{latexsym}
\usepackage{graphicx}
\title{Bent crystal selection and assembling for the LAUE project} 
\author{V.~Liccardo\supit{a,c}, E.~Virgilli\supit{a}, F.~Frontera\supit{a,b}, V.~Valsan\supit{a,c},  
\\V.~Guidi\supit{a}, E.~Buffagni \supit{d}
\skiplinehalf
\supit{a} \small\textit{Physics and Earth Sciences Department, University of Ferrara, via Saragat, 1 - 44122 Italy};\\
\supit{b} \small\textit{IASF-INAF via P.Gobetti, Bologna - Italy};\\
\supit{c} \small\textit{Universit\'e de Nice Sophia-Antipolis, Parc Valrose, 06108 Nice Cedex 2, France};\\
\supit{d} \small\textit{IMEM - Parco Area delle Scienze 37/A - 43124 Parma, Italy}.\\
}

\authorinfo{E-mail to: liccardo@fe.infn.it, virgilli@fe.infn.it}

\begin{document} 
\maketitle

\begin{abstract}
For the first time, with the Laue project, bent crystals are being used for focusing photons in the 
80-300 keV energy range.  The advantage is their high reflectivity and better Point Spread Function with 
respect to the mosaic flat crystals. Simulations have already shown their excellent focusing capability 
which makes them the best candidates for a Laue lens whose sensitivity is also driven by the size of the 
focused spot. Selected crystals are Germanium (perfect, (111)) and Gallium Arsenide (mosaic, (220)) with 
40 m curvature radius to get a  spherical lens with 20 m long focal length. A lens petal is being built. We report 
the measurement technique by which we are able to estimate the exact curvature of each tile within a few 
percent of uncertainty and their diffraction efficiency. We also discuss some results.
\end{abstract}


\section{INTRODUCTION}
\label{sec:intro}

In the framework of the Laue project [\citenum{Frontera13}], [\citenum{Virgilli11b}], one of the main 
goals is the development of a technology for the production of bent crystals with the required 
characteristics, like proper curvature, high efficiency, low mosaicity. The advantage of curved 
crystals is their high diffraction efficiency and their capability of better concentrating the 
signal collected over a large area into a small focal spot [\citenum{Barriere08}], [\citenum{Frontera10}].
This challenging goal is being achieved, thanks to the cooperation of many institutions. Thus, for the 
first time a Laue lens petal completely made of bent crystals is being assembled.  

A previous set of measurements [\citenum{Liccardo12}], performed in the LARIX facility of the University 
of Ferrara, has shown that bent mosaic crystals of GaAs, diffractive planes (220), and bent perfect crystals of Ge (111) are 
very good candidates. They are resulted to show a good reproducibility in terms of curvature and efficiency. 
Thus they have been adopted for assembling a lens petal. In this paper, we will report the method used to 
measure the curvature of these bent crystals and the curvature measurement results.

\section{Facility set-up} 
\label{sec:facility}

\begin{figure}[!h]
  \begin{center}
  \includegraphics[width=0.6\hsize, angle=0]{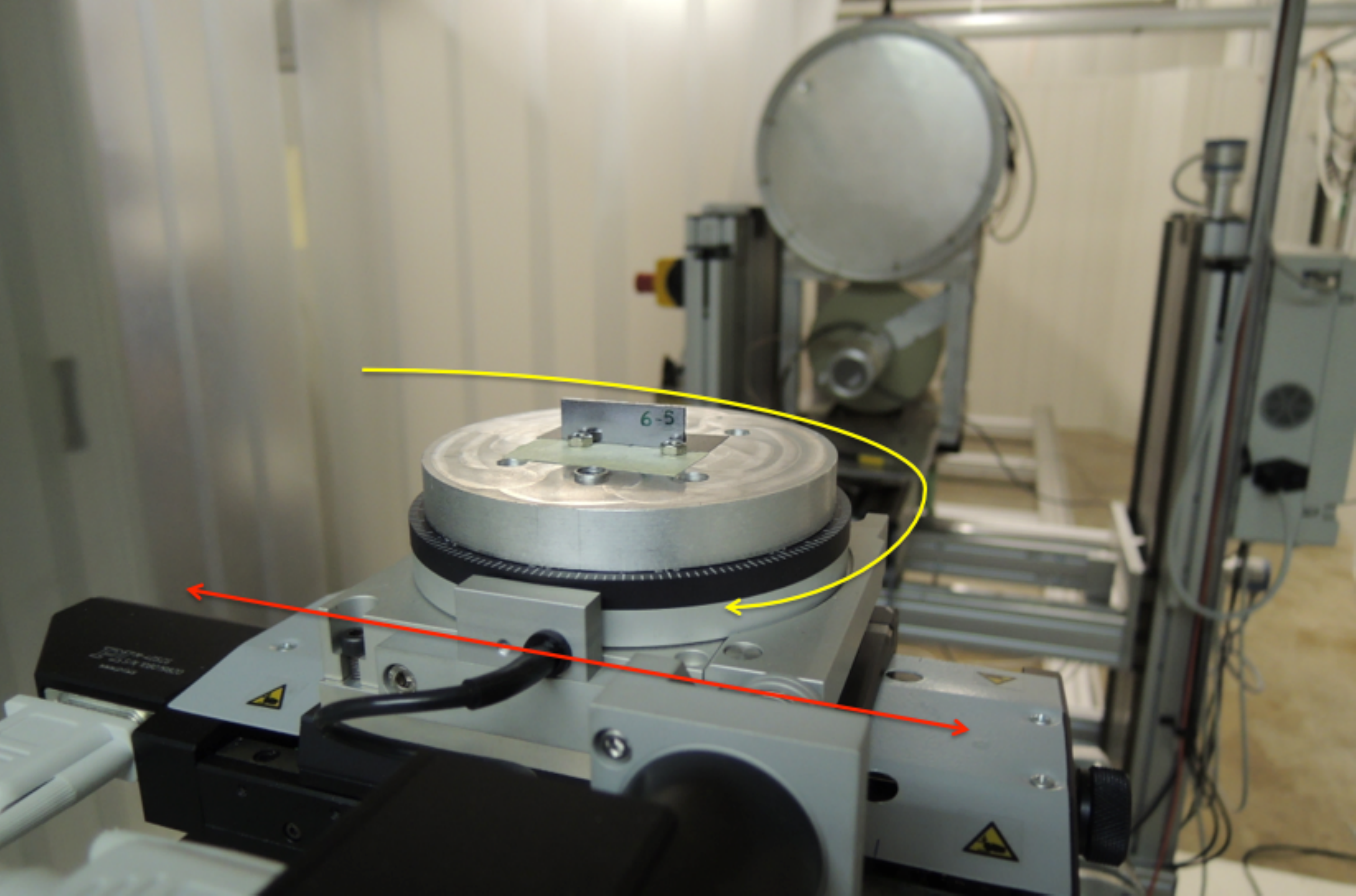}
  \caption{\footnotesize{The new platform with the upgraded motors.}}\label{stage}
\end{center}
\end{figure}

The measurements have been carried out in the LARIX facility of the University of Ferrara. 
A detailed description of the set up adopted can be found in Ref. [\citenum{Liccardo12}]. 
However, some changes have been made since the last set of measurements to get more accurate 
results. In particular, the translation and rotation stages of the crystal holder have been 
upgraded, achieving an accuracy of 1 $\mu$m along a direction perpendicular to the incident 
beam, and of 3 arcsec for the Bragg angle determination (see Fig.~\ref{stage}). These movements 
are monitored with an external system of optical encoders. In this way the true position of
 the sample and the position of the XYZ holder (X being the beam direction) can be accurately 
measured. The two beam collimators have been adjusted in such a way to provide a beam size 
of 0.5 mm $\times$ 3 mm and a divergence value of about 18 arcsec. The first collimator does 
not significantly influence the size of the beam but affects the beam intensity, whereas the 
second collimator is crucial to get the size of the spot needed for the specific application.
In the current configuration, the setup can be used for measuring the crystal curvature, and the 
beam photon angular spread, from which we can determine the crystal mosaicity.

\section{Crystal bending} 
\label{sec:diffraction}

The bent crystals of Ge (111) are provided by the Sensor and Semiconductor Laboratory (LSS) of the University of 
Ferrara. The bending is obtained by means of a series of parallel superficial indentations on one of the largest faces of 
the crystals [\citenum{Camattari11}], [\citenum{Guidi12}], resulting in a primary curvature perpendicular to the indentation 
and a secondary curvature, also called quasi-mosaic, which acts with the crystal, perpendicular to the main curvature
[\citenum{Guidi11}].  

Bent mosaic crystals of GaAs (220) are instead provided by CNR-Institute of Materials for Electronics 
and Magnetism (IMEM) of Parma, Italy. The crystal curvature is obtained by a controlled surface damaging, 
which introduces defects in a superficial layer of few tens of nanometers in thickness undergoing a highly 
compressive strain. The controlled surface damaging were obtained by means of a mechanical lapping process on 
one of the two sides of the planar samples [\citenum{Buffagni11}]. 

\begin{figure}[!h]
  \begin{center}
  \includegraphics[width=0.4\hsize, angle=0]{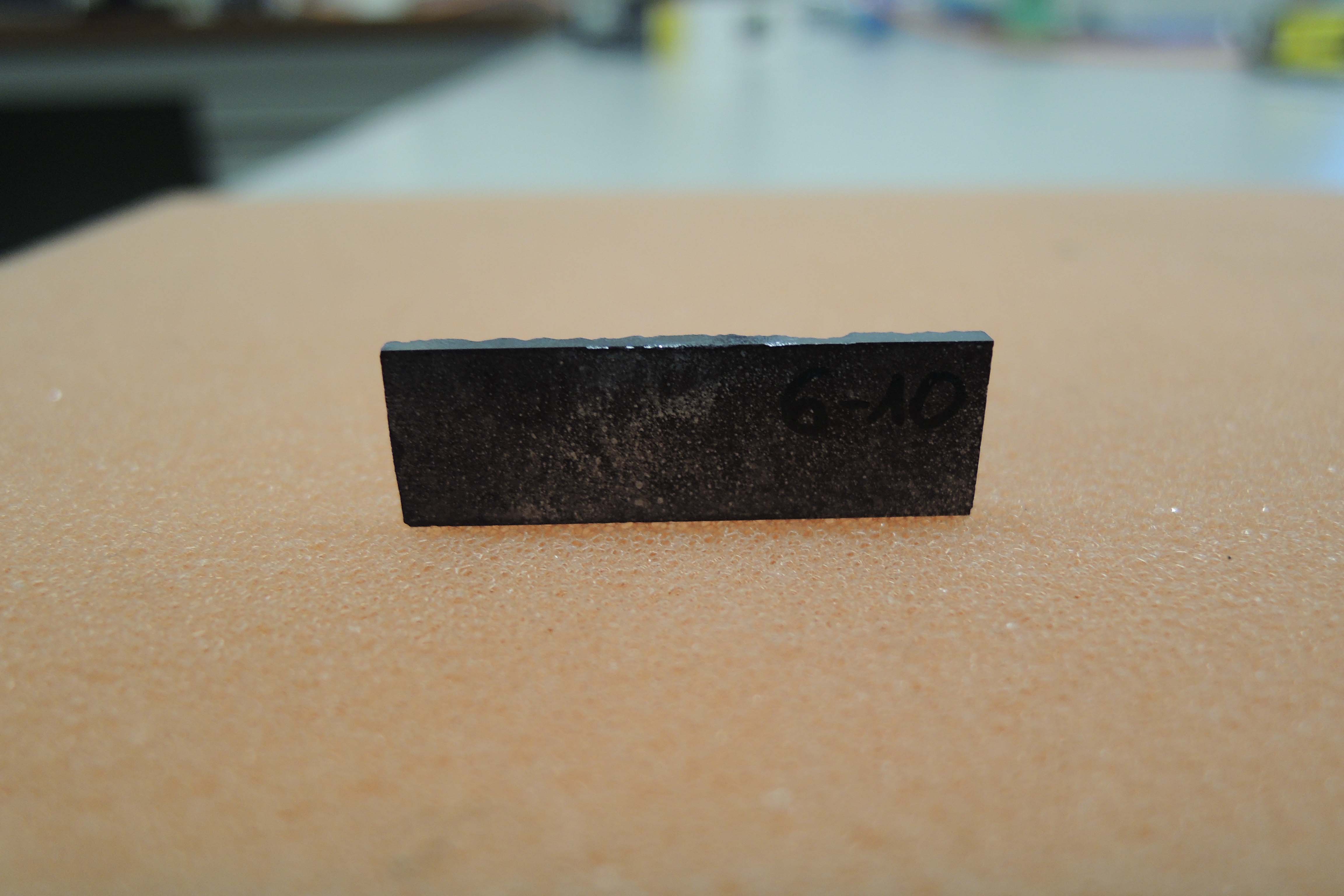}
  \includegraphics[width=0.4\hsize, angle=0]{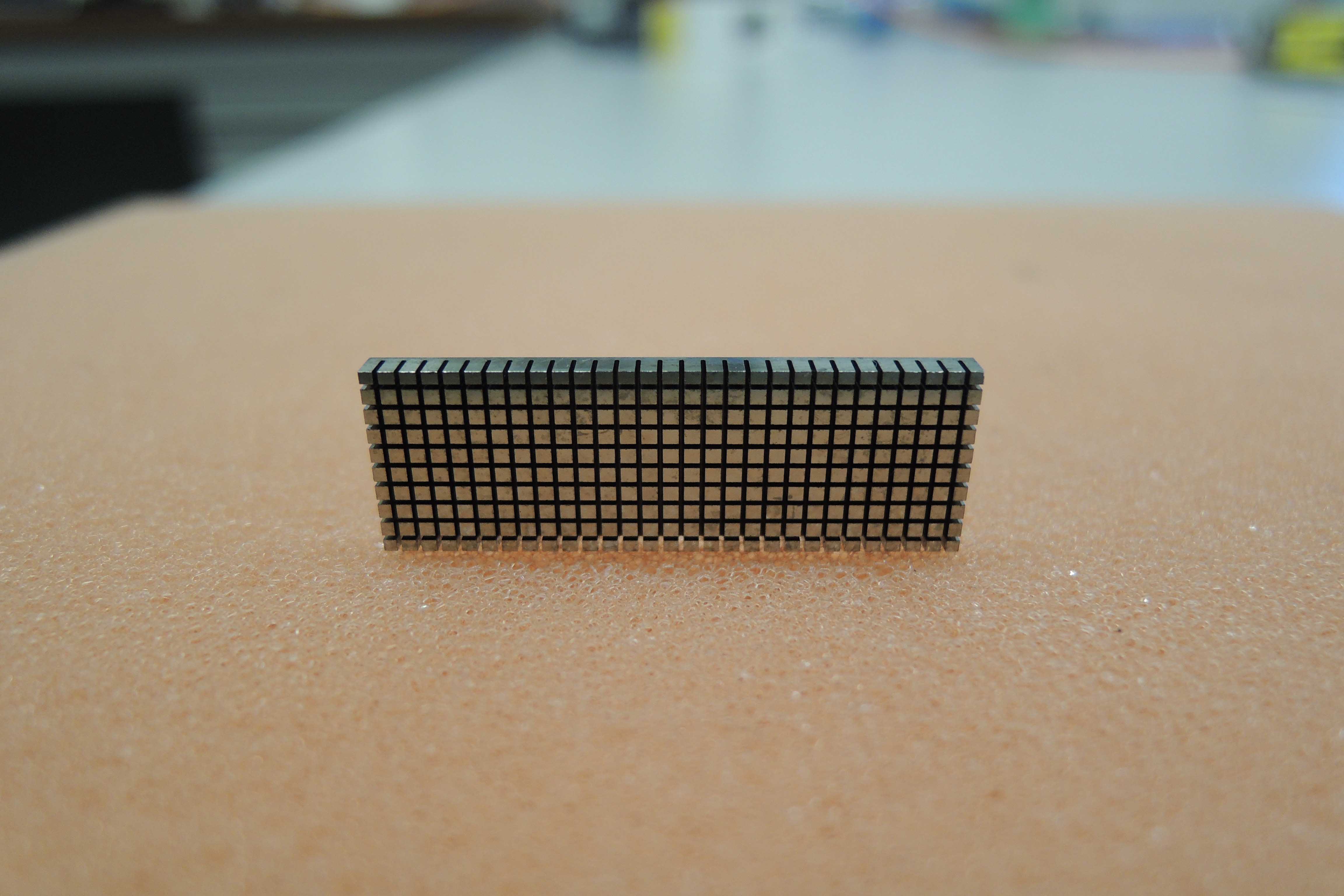}
  \caption{\footnotesize{\textit{Left}: Gaas (220) crystal tile. \textit{Right}: Ge (111) crystal tile }}\label{gaas}
\end{center}
\end{figure}

With the developed setup, it is possible to analyze the curvatures of both set of crystals. The required crystal 
tile size of both crystal types is 30 mm $\times$ 10 mm, with a thickness of 2 mm (Fig.~\ref{gaas}).

For the lens petal prototype, which is being assembled, LSS has already provided all the bent Ge (111) crystal 
tiles (about 150) with a nominal angular spread of 4 arcsec, but only 39 crystals have been analysed in order to test 
the curvature obtained by means of the indentations. The  curvature is preliminarly measured by the LSS staff 
means of a profilometer, obtaining a curvature radius of 40m.  However, it is important to take into account 
that the measured curvature just estimates the profile of the external surface of the single crystals. 
For this reason we have carried out, in the LARIX facility, diffraction tests in transmission geometry, to 
better investigate the curvature of the lattice planes, obtaining some deviations with respect the profilometer 
results. The indented Ge samples have an excellent regularity in thickness (2$\div$2.01 mm) and weight (2.06$\div$2.08 g).

IMEM has supplied 2 sets of bent crystal tiles of GaAs (220). The first set was composed of 10 samples of 
GaAs (220) with variable thickness (0.45$\div$0.55 mm) and 
square cross section (15 mm $\times$ 15 mm), whose curvature was measured by themselves with a Bragg 
diffractometer, that makes use of  a Cu K$\alpha$ fluorescence line. This set was developed to compare 
the results obtained by them with those obtainable with the LARIX facility. 
The samples coming from the same ingot resulted to exhibit a uniform mosaicity (mosaic-spread $\sim$ 15-18 arcsec). 
Each sample resulted to show two different curvatures along the orthogonal axes $x$ and $y$. The IMEM group was 
able to estimate the curvature radius with an accuracy of 5\%. However, in Bragg geometry,  the estimate of the 
curvature is based on the curvature of the shallow layers of the sample, as the deeper regions of the crystal 
are not irradiated by X-rays. Therefore, we decided to perform the transmission measurement in the LARIX facility in order to make a comparison with the results obtained by IMEM.
The second set of 18 GaAs (220) crystals (with mosaicity of 25 arcsec) provided by IMEM, has instead the size 
required by the project (30 mm $\times$ 10 mm $\times$ 2 mm) and a variable weight (2.40$\div$2.85 g), 
depending on the different amount of material removed by the lapping procedure.

\section{Adopted method for the curvature measurement}
\label{sec:experimental}

To estimate the value of the primary curvature (also called main curvature) of a sample, we adopted 
the following method. The sample is fixed to the moving system that allows to translate the crystal 
perpendicular to the direction of the incident beam, and rotate it around a vertical axis perpendicular 
to the beam and passing through the center of the tile itself. Then the crystal diffraction of the
 monochromatic radiation at 59.2 keV due to the  $K\alpha$ fluorescence of the X--ray tube W anode, 
is analyzed in several crystal points. From the comparison of the spectra of the diffracted beams from 
each point it is possible to estimate the curvature. 

\begin{figure}[!h]
  \begin{center}
  \includegraphics[width=0.5\hsize, angle=0]{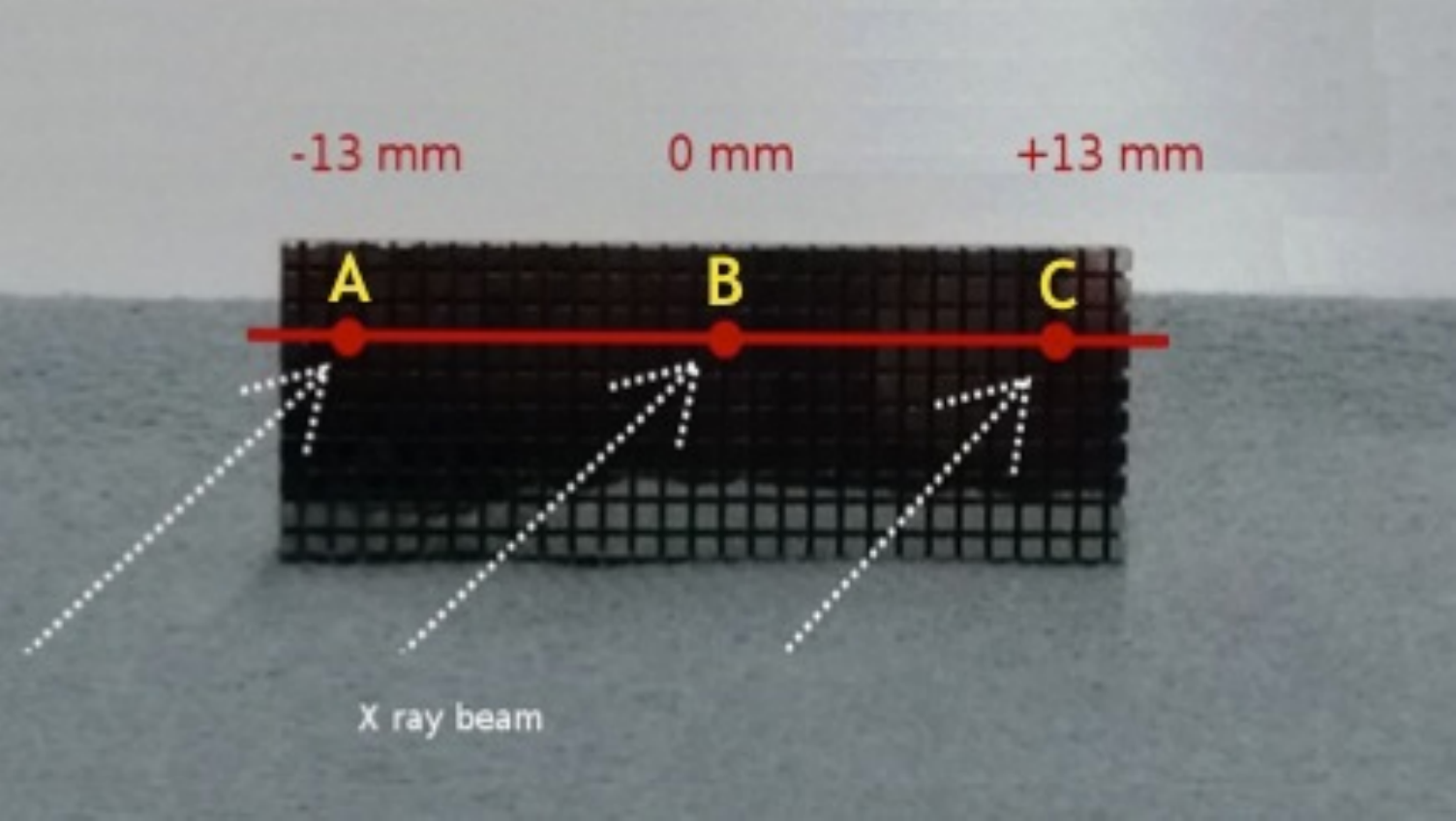}
  \caption{\footnotesize{The sample is analyzed along several points lying on the same 
horizontal plane. The same test is performed by raising and lowering the sample, 
to investigate different areas of the crystal.}}
\label{abc}
\end{center}
\end{figure}

In particular, the first step is to determine the Bragg angle $\alpha$, in the point A of the crystal 
(Fig.~\ref{abc}). The crystal is then moved parallel to itself so that the beam impinges on the crystal 
in a different point B, with distance AB well known. Since the tile is bent, the point B needs a different 
angle to diffract the same energy as the point A. To get the same diffracted energy (59.2 keV) in 
correspondence of the point B, we have to rotate the sample of an angle $\Delta$$\alpha$ which 
represents the angular aperture between the points A and B. Applying the same procedure to a set 
of points along the crystal, it is possible to investigate the angular aperture with continuity 
along the tile, obtaining the mean curvature radius of the crystal itself.

The estimation of the curvature radius  is made by performing a linear fit to the different rotation 
angles measured. Indeed the angular coefficient of the straight line gives the curvature of the 
sample (Fig.~\ref{ge111}). The uncertainty on the curvature radius is related to the error in the 
determination of the distance between the contiguous points where the energy of the diffracted beams 
are measured, and on the determination of the Bragg angle corresponding to the $K{\alpha}$ line as 
well. Being $\Delta$$l$ the distance between the points, and $\Delta$$\theta$ the difference between 
the Bragg angles of the points diffracting the same energy, the curvature radius $R$ and its 
uncertainty $\Delta R$ are given by:

\begin{equation}\label{radius}
R = \frac{{\Delta l}}{{\Delta \theta }}, ~~~~~~ \Delta R = R\sqrt {{{\left( {\frac{{{\sigma _l}}}{{\Delta l}}} \right)}^2} + {{\left( {\frac{{{\sigma _\theta }}}{{\Delta \theta }}} \right)}^2}}
\end{equation}

The uncertainty $\sigma_l$ in the estimation of the distance between the two 
points is related to the accuracy of the translational stage which moves the 
crystal and allows the beam to impinge on two adjacent points of the sample. As written 
above, this accuracy is 0.001 mm, whereas the typical distance between two successive 
points is 4-5 mm. It follows that: 

\begin{equation}\label{uncertainty}
\left( {\frac{{{\sigma _l}}}{{\Delta l}}} \right) \sim 2.5 \times {10^{ - 5}}
\end{equation}

Regarding the angular uncertainty, the actuator used has an accuracy of $10^{-4}$ degrees but the Bragg angle
 estimation necessary to diffract the desired energy is determined by a Gaussian fit of the Rocking curve, 
which reduces the value of the error to 2 $\times$ 10$^{-5}$ degrees. The variation of the angle between two adjacent 
points, being dependent on the distance between the points and the curvature to be measured (approximately 40 m), 
is of the order of 0.02$^{\circ}$:

\begin{equation}\label{uncertainty1}
\left( {\frac{{{\sigma _\theta }}}{{\Delta \theta }}} \right) \sim 0.025
\end{equation}

It is evident that the main error comes from the angular uncertainty, while the translational 
stage error is negligible. The accuracy in determining the angle variation obviously increases 
if the points to be compared are farther from each other. In this case the difference $\Delta$$\theta$ 
between the Bragg angles relative to the points analyzed is greater, whereas the uncertainty in the 
determination of each point is the same, improving the estimation of the curvature radius. Based on 
the above considerations we can estimate the curvature radius  with an error of about 2 m, for those 
samples which exhibit a bending radius of 40 meters.

\section{Curvature measurement Results}

\subsection{Germanium (111) crystal tiles}
\label{sec:germanium}

\begin{figure}[!h]
  \begin{center}
  \includegraphics[width=0.45\hsize, angle=0]{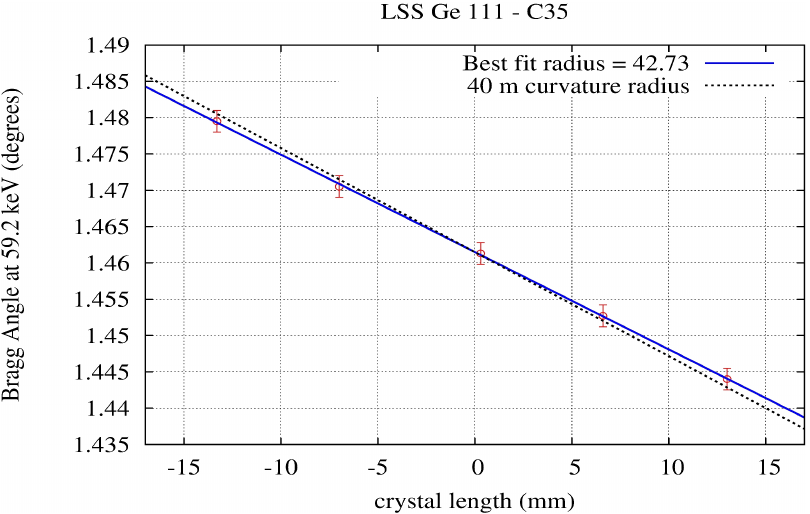}
  \includegraphics[width=0.45\hsize, angle=0]{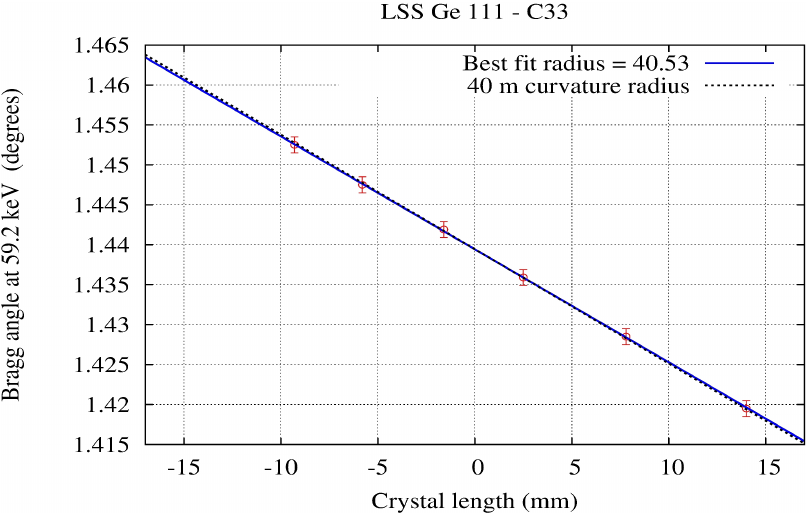}
  \caption{\footnotesize{Bending estimations for two samples of Germanium (C35 and C33). 
The measured radii are compared with the expected curvature (40 m). The $x$ and $y$ axes 
show respectively the distance from the center of the crystal (zero), and the corresponding 
59.2 keV Bragg angle.}}
\label{ge111}
\end{center}
\end{figure}

All the Ge (111) crystal tiles show a bending radius extremely uniform throughout 
the crystal, even if in some case there is a discrepancy between the expected and 
the measured curvature.
As regards those samples characterized by a curvature radius greater than expected, 
it has been experimentally demonstrated in a first stage, that by indenting again the 
tile increasing the grooves depth, the curvature radius decreases. 
On the other hand, if the curvature radius is smaller than expected, the crystals can 
be further exposed to an etching procedure by means it is possible to increase the 
curvature radius of the tile. 
Therefore the crystal profiles can be adjusted, with an accuracy of 5\%, once the 
diffraction tests have been performed. Two examples of crystal curvature measurement 
method are given in Fig.~\ref{ge111}.

\begin{figure}[!h]
  \begin{center}
  \includegraphics[width=0.55\hsize, angle=0]{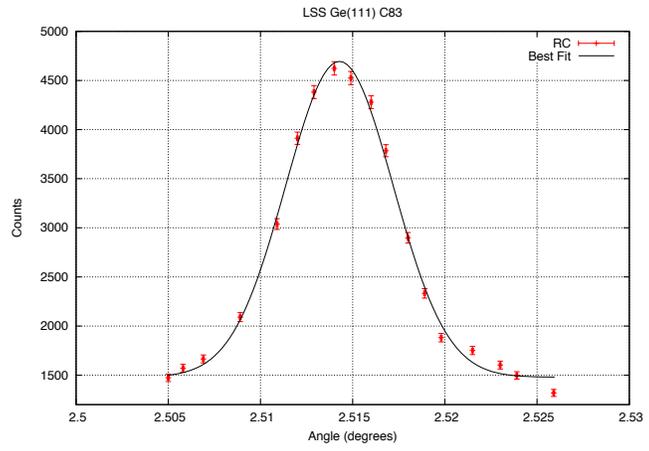}
  \caption{\footnotesize{Bent Ge (111) rocking curve as recorded at the 
LARIX facility. The FWHM of the gaussian fit turned out to be 24.6 arcsec. 
The broadening of the response function (FWHM 4 arcsec) is mainly due to the 
beam divergence (18 arcsec), which is also responsible of the gaussian shape.}}
\label{ge_rc}
\end{center}
\end{figure}

\begin{table}[!h]
\begin{center}
\resizebox{8cm}{!}{
\begin{tabular}{c c | c c | c c}
\toprule
 Crystal & Curvature & Crystal & Curvature & Crystal & Curvature\\
 Number  &  Radius (m) & Number  &  Radius (m) & Number  &  Radius (m)\\
 \midrule
8  &  44.7 & 29 & 43.0 & 58 & 41.8\\
9  & 44.2 & 30 & 40.5 & 68 & 42.8\\
10 & 37.4 & 31 & 37.4 & 71 & 43.7\\
11 &  44.3 & 32 & 26.0 & 72 & 44.7\\
13 &  44.7 & 33 & 40.5 & 73 & 43.1\\
14 &  37.3 & 34 & 43.3 & 80 & 37.9\\
15 &  40.5 & 35 & 42.7 & 81 & 38.4\\
16 &  38.5 & 36 & 38.0 & 82 & 44.2\\
17 & 41.8 & 37 & 33.0 & 83 & 30.4\\
21 & 39.5 & 47 & 54.5 & 92 & 40.9\\
24 &  46.4 & 48 & 45.7 & 94 & 52.6\\
25 & 51.1 & 49 & 39.4 & 94 & 52.6\\
26 &  47.7 & 50 & 41.3 & 119 & 39.3\\
27 & 43.5 & 51 & 42.3 & 126 & 47.0\\
28 & 41.5 & 56 & 41.2 & 153 & 44.3\\
\bottomrule
\end{tabular}
}
\end{center}
\caption{\footnotesize Ge(111) bending radii of the samples analyzed for the LAUE project.}
\label{tab:geradii}
\end{table}

\subsection{Gallium Arsenide (220) crystal tiles}
\label{sec:gallium}

The first set of crystals supplied by IMEM have shown a good agreement between 
the results obtained using the reflection diffractometer and transmission technique 
adopted in the LARIX facility. Table~\ref{tab:gaasradii} reports the results obtained by IMEM 
compared with those obtained in the LARIX. With the exception of the curvature along the 
$y$ axis measured for the crystal number 3 (whose bending radius is very small, therefore 
unsuitable for the project) for which we obtain a discrepancy between the two measures 
greater than 10\%, the measured curvature radii, along both $x$ and $y$ axes, agree with 
each other within an uncertainty of 
5\%.

\begin{table}[!h]
\begin{center}
\resizebox{12cm}{!}{
\begin{tabular}{c c c c | c c | c c | c}
\toprule
 Crystal & Thickness & Grain-paper &  \multicolumn{2}{c}{IMEM} & \multicolumn{2}{c}{LARIX} & \multicolumn{2}{c}{Deviation (\%)} \\ \cmidrule(r){4-5} \cmidrule(r){6-7}  
 & & & Rc $x$  & Rc $y$ & Rc $x$ & Rc $y$ &  \multicolumn{2}{c}{IMEM-LARIX}\\ \cmidrule(r){8-9} 
 Tile  &  (mm) &  & (m) &  (m) &  (m)  &  (m) &  x-axis & y-axis\\
 \midrule
1  & 0.495 & P2500 & 13.8 & 66.7 & 14.3 & 67.9 & 3.5\% & 1.8\%\\
2  & 0.440 & P600 & 6.0 & 23.3 & n.t. & 22.8  & - & 2.1\%\\
3 & 0.460 & P320 & 5.2 & 16.2 & 5.8 & n.t. & 10.9\% & -\\
4 & 0.502 & P4000 & 20.3 & 60.0 & 20.6 & 59.6 & 1.5\% & 0.7\%\\
5 & 0.420 & P180 & 3.1 & 7.4 & n.t. & n.t. & - & -\\
6 & 0.445 & P1200 & 7.4 & 29.2 & n.t. & 30.3 & - & 3.6\%\\
7 & 0.435 & P400 & 5.6 &10.6 & n.t. & 11.0 & - & 4.1\%\\
\bottomrule
\end{tabular}
}
\end{center}
\caption{\footnotesize GaAs(220) bending radii of the samples analyzed 
for the LAUE project (n.t. corresponds to "not tested").}
\label{tab:gaasradii}
\end{table}

\begin{figure}[!h]
  \begin{center}
  \includegraphics[width=0.45\hsize, angle=0]{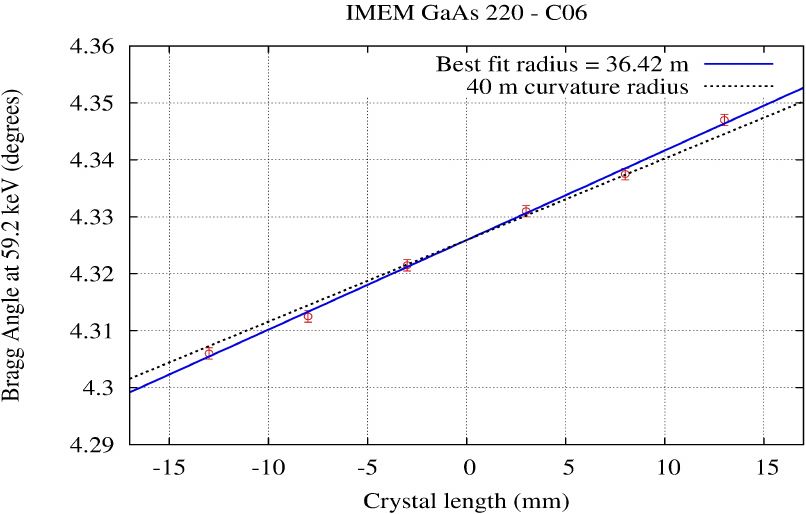}
  \includegraphics[width=0.45\hsize, angle=0]{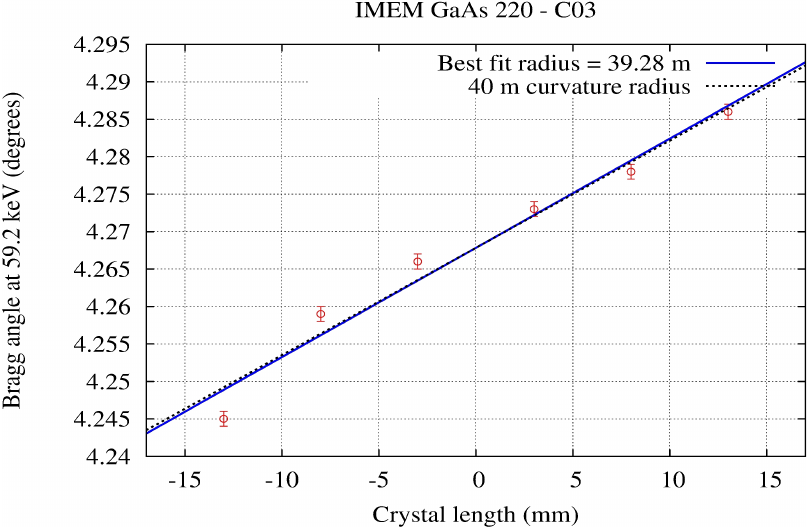}
  \caption{\footnotesize{Bending estimations for two samples of GaAs (C06 e C03). 
The measured radii are compared with the expected curvature (40 m). The $x$ and $y$ axes show 
respectively the distance from the center of the crystal (zero), and the corresponding 
59.2 keV Bragg angle.}}
\label{gaas220}
\end{center}
\end{figure}

\begin{figure}[!h]
  \begin{center}
  \includegraphics[width=0.55\hsize, angle=0]{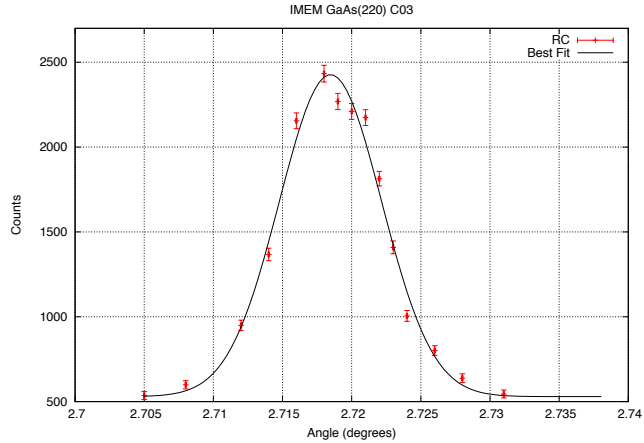}
  \caption{\footnotesize{Mosaic GaAs (220) rocking curve as recorded at the 
LARIX facility. The FWHM (31.2 arcsec) of the gaussian fit in this case, as we expect, 
is greater than Ge (111) crystals. After the correction of the beam divergence, the
mosaicity value obtained was 25.5 arcsec.}}
\label{gaas_rc}
\end{center}
\end{figure}

The second set of crystals provided by IMEM (see Sec.~\ref{sec:diffraction}) satisfies the curvature required
 by the LAUE project. The results obtained at IMEM and LARIX laboratory are reported in Table~\ref{tab:gaasradii2}. 
An example of rocking curve, which is used to estimate the crystal mosaicity, is shown in  Fig.~\ref{gaas_rc}. 
The discrepancy between the LARIX and IMEM estimates in this case is greater. This discrepancy likely is 
due to the fact that IMEM measures the reflected beam, while we measure the diffracted beam in transmission 
geometry, and also to the non-uniform curvature of the GaAs samples due to the presence of dislocations. Furthermore, a non-uniformity for the GaAs samples 
is observed at the edges of the tiles, if compared to the Germanium samples (Fig.~\ref{gaas220}).

\begin{table}[!h]
\begin{center}
\resizebox{10cm}{!}{
\begin{tabular}{c c c c c c}
\toprule
 Crystal & Weight & Rc & Rc & Deviation (\%) & Mosaicity (arcsec)\\
 Number  & (g) & IMEM  &  LARIX & IMEM-LARIX  &   LARIX\\
 \midrule
1  &  2.40 & 40.3 & 35.0 & 13.9\% & 25.5\\
2  & 2.05 & 42.4 & 39.5 & 7.3\% & 27.4\\
3 & 2.66 & 41.9 & 39.3 & 6.5\% & 23.6\\
4 &  2.67 & 38.8 & 39.2 & 1.1\% & 21.2\\
4.8 & 2.55 & 39.4 & 35.9 & 8.75\% & -\\
4.9 & 2.62 & 38.7 & 36.3 & 6\% & -\\
5 & 2.85 & 40.0 & 38.2 & 4.5\% & 25.2\\
5.2 & 2.35 & 40.0 & 39.9 & 0.2\% & -\\
5.10 & 2.42 & 40.0 & 40.0 & -\% & -\\
6 & 2.60 & 40.0 & 36.5 & 9.2\% & 23.4\\
6.4 & 2.65 & 40.0 & 38.9 & 0.2\% & -\\
6.6 & 2.62 & 42.0 & 40.5 & 3.75\% & -\\
6.10 & 2.52 & 38.8 & 36.8 & 5.0\% & -\\
6.11 & 2.60 & 38.6 & 34.7 & 9.7\% & -\\
6.12 & 2.55 & 41.1 & 42.38 & 3.2\% & -\\
\bottomrule
\end{tabular}
}
\end{center}
\caption{\footnotesize Bending radii of the second set of GaAs (220) samples provided by IMEM.}
\label{tab:gaasradii2}
\end{table}

\section{Discussion and Conclusions}
\label{sec:discussion}

The performed curvature tests have shown that the production of the curvature of 
the crystal tiles for the lens petal prototype satisfies the requirements within 5-10\% accuracy. 

\begin{figure}[!h]
  \begin{center}
  \includegraphics[width=0.35\hsize, angle=0]{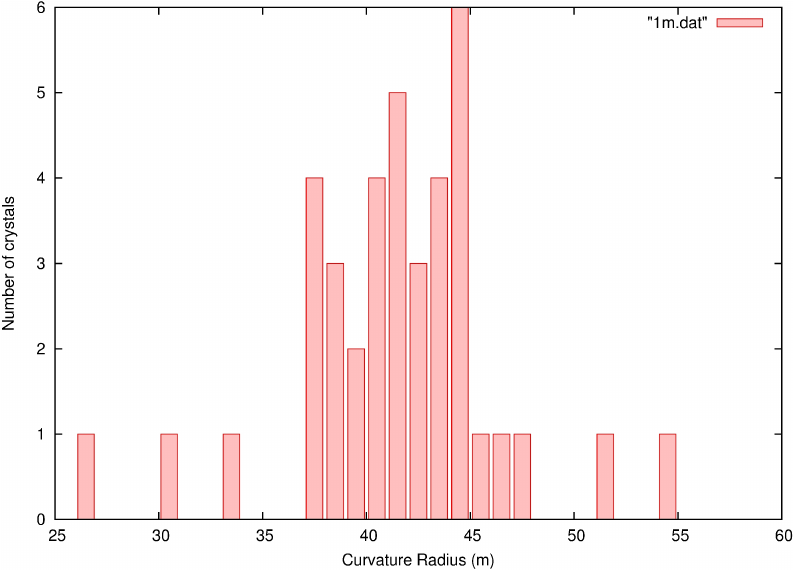}
  \includegraphics[width=0.38\hsize, angle=0]{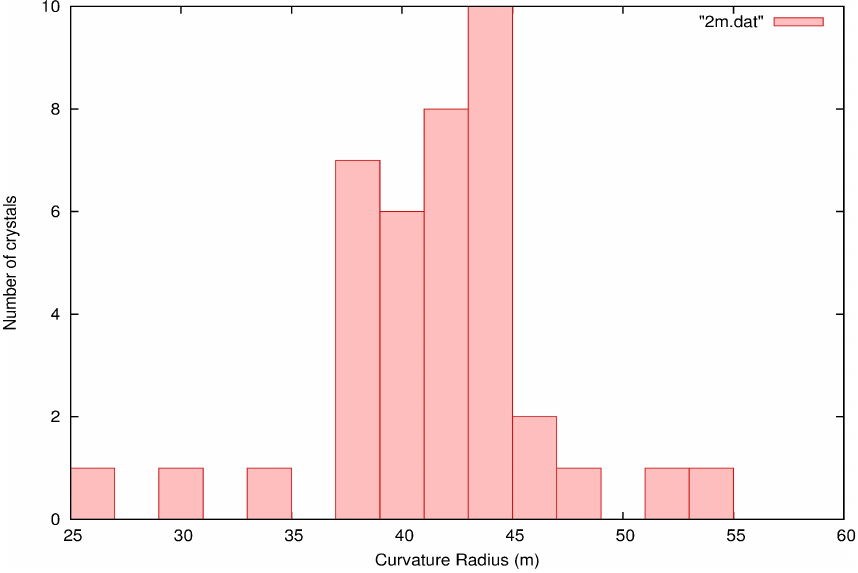}
  \caption{\footnotesize{Distribution of the bending radii of the 39 
Ge (111) samples analyzed, with a rebinning of 1m (\textit{Left}) 
and 2 m (\textit{Right}).}}
\label{stat}
\end{center}
\end{figure}

In particular, concerning Ge (111) crystal tiles, 36\% of the tested tiles have a curvature 
radius of 40~m within 5\%, while 60\% of the tiles have a curvature radius of 40~m within 10\%. 
In Figure \ref{stat} it is shown the distribution of the measured curvature radii. 

\begin{figure}[!h]
  \begin{center}
  \includegraphics[width=0.4\hsize, angle=0]{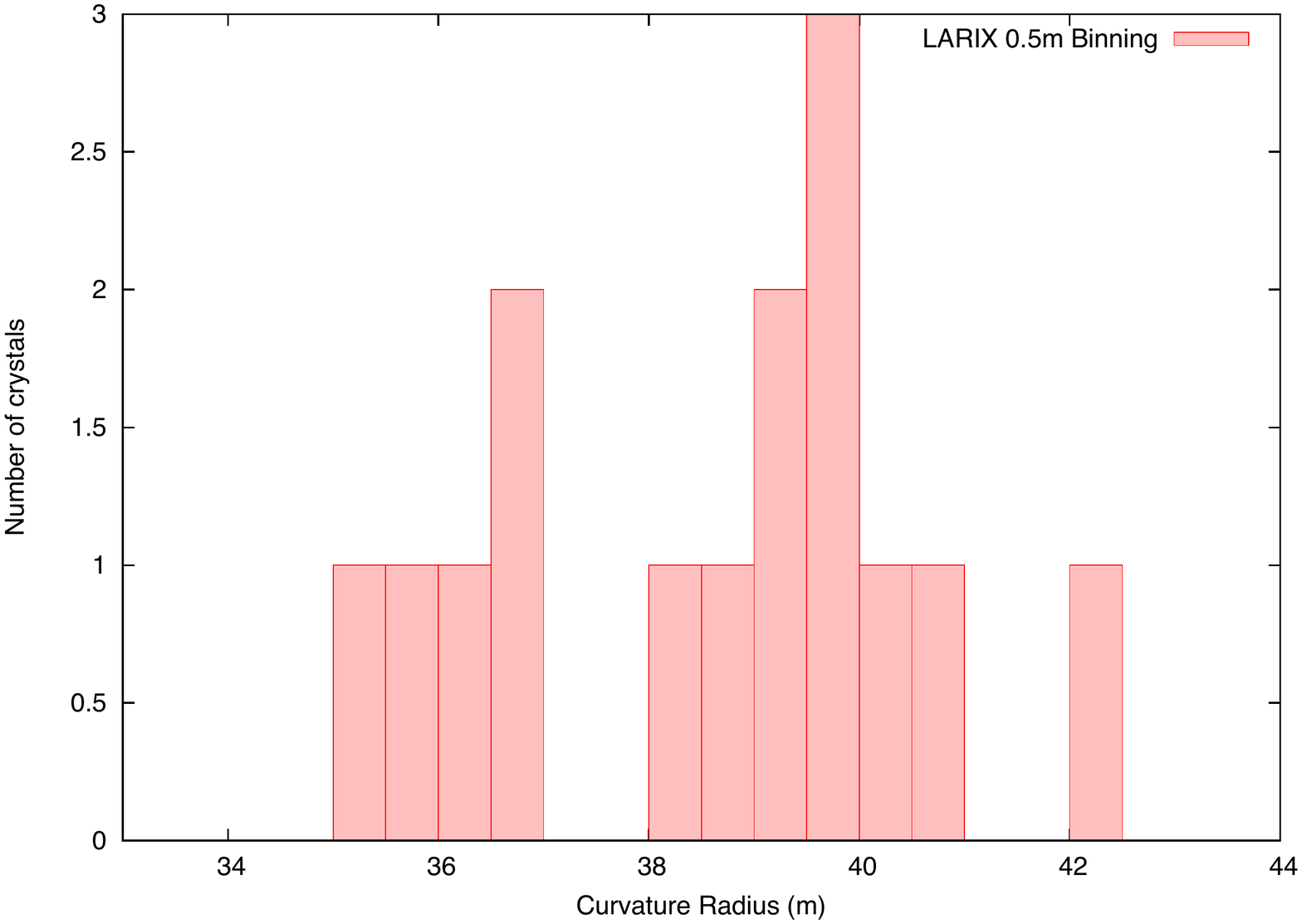}
  \includegraphics[width=0.4\hsize, angle=0]{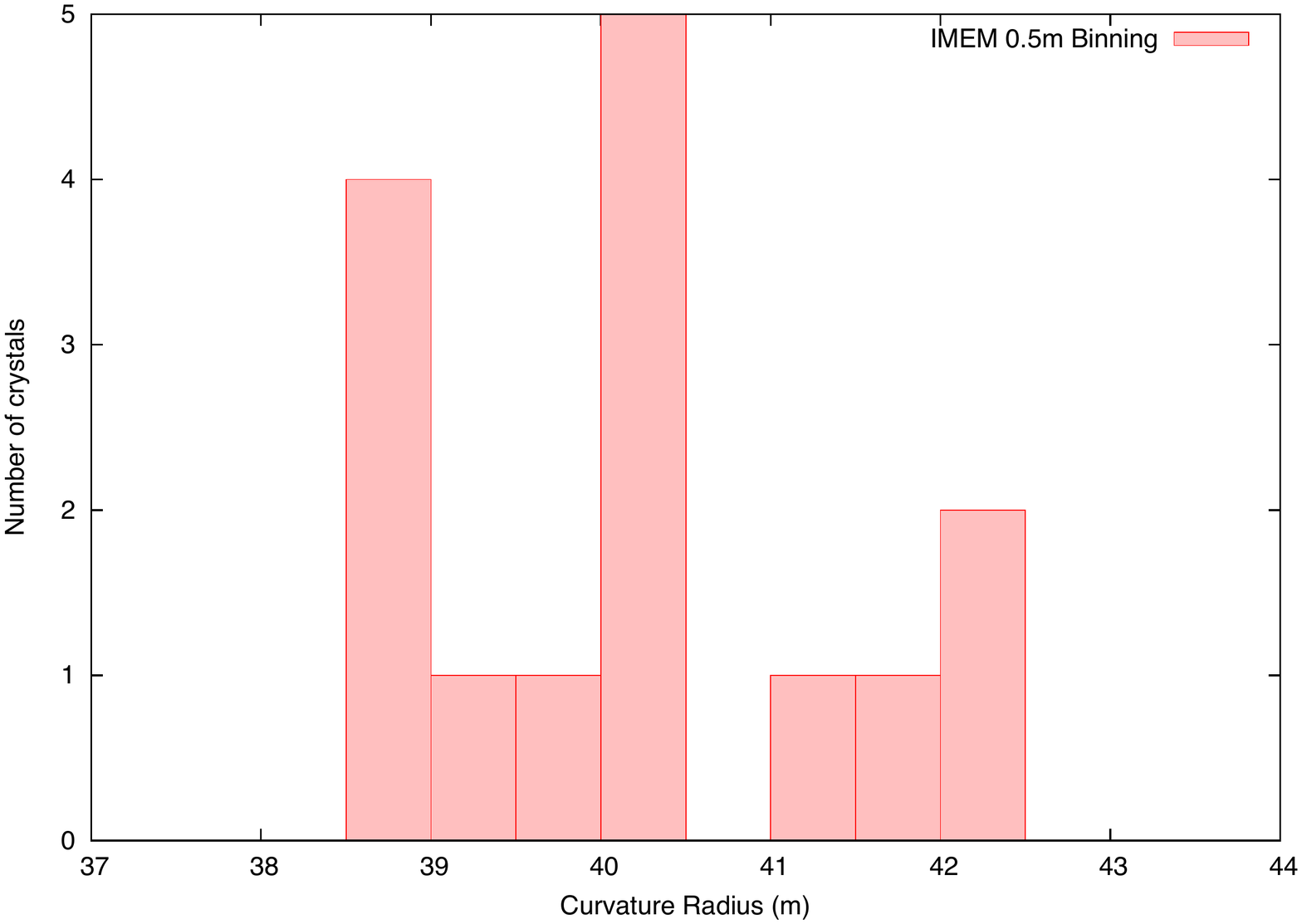}
  \caption{\footnotesize{\textit{Left}: Distribution of the bending radii of 15 GaAs (220) 
samples as estimated at LARIX facility. \textit{Right}: Distribution of 
radii as obtained at IMEM Parma.}}
\label{stat1}
\end{center}
\end{figure}

Concerning GaAs (220) crystal tiles, due to the small number of tested crystals it 
has not been possible to make a similar statistical analysis to make an inference 
about the goodness of the tiles provided (see Fig.~\ref{stat1}). However, from the results obtained
we can see that 44\% of the samples tested have a curvature of 40 m ($\pm$5\%), whereas 83\% 
($\pm$10\%).

\begin{figure}[!h]
  \begin{center}
  \includegraphics[width=0.45\hsize, angle=0]{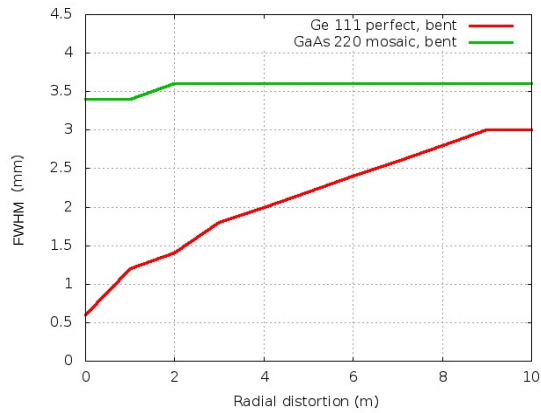}
  \caption{\footnotesize{PSF size (FWHM) of an entire Laue lens (focal length of 20 m) having 
a 50-600 keV passband, composed of Ge (111) (red curve) or GaAs (220) (green curve) crystals as 
a function of the deviation of the crystal curvature from the 40 m bending radius.}}
\label{misal}
\end{center}
\end{figure}

Together with the measurements of the crystal samples, a series of simulations were 
performed [\citenum{Valsan13}] in order to determine the maximum acceptable radial 
distortion with respect to the nominal curvature radius. In 
Figure~\ref{misal} it is shown the  dependence of the Full Width at Half Maximum (FWHM) 
of the Point Spread Function (PSF) on the crystal deviation from the nominal curvature 
(40 m), for a lens entirely made of either Ge (111) (red curve) or GaAs (220) (green curve) 
crystals. The simulation shows that for a distortion equal to zero (the ideal case), 
the FWHM is equal to 0.6 mm in the case of Ge perfect crystals, whereas it is equal to 
3.4 mm using mosaic GaAs (220) tiles. Despite the Germanium PSF rapidly grows with the 
increasing distortion, it is always smaller than that of GaAs whose size seems to be 
almost constant. Therefore a small distortion is acceptable for a lens made of Ge (111) crystals.

\acknowledgments     
 
The authors wish acknowledge the financial 
support by the Italian Space Agency (ASI) through the project ``LAUE - Una Lente
per i raggi Gamma''under contract I/068/09/0.
V.V. and V.L. are supported by the Erasmus Mundus Joint Doctorate Program by 
Grant Number 2010-1816 from the EACEA of the European Commission.

\bibliography{crystal_tests}
\bibliographystyle{spiebib}

\end{document}